\documentclass[twocolumn,prl,amsfonts]{revtex4}
\usepackage{amsmath}
\usepackage{bm}
\newcommand{\ket}[1]{\ensuremath{|#1\rangle}}
\newcommand{\bra}[1]{\ensuremath{\langle#1|}}
\newcommand{\nuc}[2]{\mbox{${}^{#1}\rm #2$}}
\newcommand{\half}{\mbox{$\frac{1}{2}$}}
\newcommand{\smhalf}{\raisebox{0.4ex}{$\scriptstyle\frac{1}{2}$}}
\newcommand{\NOT}{\textsc{not}}
\newcommand{\CNOT}{controlled-\NOT}
\newcommand{\SWAP}{\textsc{swap}}
\begin{document}
\title{NMR Quantum Computation: a Critical Evaluation}
\author{J. A. Jones}
\affiliation{Centre for Quantum Computation, Clarendon Laboratory,
Parks Road, Oxford OX1 3PU, UK}\affiliation{ Oxford Centre for
Molecular Sciences, New Chemistry Laboratory, South Parks Road,
Oxford OX1 3QT, UK}
\begin{abstract}
Liquid state nuclear magnetic resonance (NMR) techniques have
produced some spectacular successes in the construction of small
quantum computers, and NMR is currently by far the leading
technology for quantum computation. There are, however, a number
of significant problems with any attempt to scale up the
technology to produce computers of any useful size.  While it is
probable that some of these will be successfully sidestepped
during the next few years, it is unlikely that they will all be
solved; thus current liquid state NMR techniques are unlikely to
provide a viable technology for practical quantum computation.
\end{abstract}
\maketitle

\section{Introduction}
Nuclear Magnetic Resonance (NMR) \cite{Ernst:1987, Goldman:1988,
Munowitz:1988, Emsley:1992, Hore:1995, Freeman:1997} is almost
unique among potential quantum technologies in that it has already
been used to build small quantum computers \cite{Cory:1996,
Cory:1997, Gershenfeld:1997, Jones:1998a, Jones:2000b}. Although
other techniques have been used to implement quantum logic gates,
such as the ion trap \CNOT\ gate \cite{Monroe:1995}, NMR provided
the first complete implementation of a quantum algorithm
\cite{Jones:1998a}.  Since then progress has been extremely rapid,
with demonstrations of Deutsch's algorithm \cite{Jones:1998a,
Chuang:1998a}, Grover's quantum search \cite{Chuang:1998b,
Jones:1998b, Jones:1998c}, versions of the Deutsch-Jozsa algorithm
with three \cite{Linden:1998a} and five \cite{Marx:1999} qubits,
and quantum counting \cite{Jones:1999a}.

In addition to being one of the most successful quantum computing
technologies, liquid state NMR is also among the oldest.  Although
explicit experimental demonstrations of NMR quantum computation
only date from 1996 \cite{Cory:1996}, many ``conventional'' NMR
experiments, such as COSY \cite{Jeener:1971, Aue:1976} and INEPT
\cite{Morris:1979} can in retrospect be reinterpreted as quantum
computations.  These and related experiments, based on coherence
transfer sequences, have been in use all over the world for
decades, and are regularly used to study complex biomolecules
containing thousands of nuclear spins \cite{Cavanagh:1996}. Even
older than these are simpler experiments such as selective
population transfer \cite{Pachler:1973}, which corresponds to the
implementation of a \CNOT\ gate, although in this case the gate
was not applied to spins in superposition states.

The rapid progress made in NMR quantum computation builds on this
pre-existing experimental sophistication: decades of experience in
manipulating nuclear spins in coherent states has inevitably
resulted in a wide range of ``tricks of the trade''.  Given this
success it may seem surprising that some early papers suggested
that this approach might be limited to quantum computers
containing about 10 qubits \cite{Warren:1997, Gershenfeld:1997b},
while many more modern estimates are similar at 10--20 qubits. The
explanation for this is simple: the highly developed nature of NMR
experiments, in comparison with many other putative quantum
technologies, means that the limits of the technique are fairly
well known and understood.  Thus unduly optimistic predictions
about the power of NMR quantum computation may be easily debunked
by experienced NMR spectroscopists.

\section{Structure and scope}
It is conventional when assessing proposals for quantum
computation technologies to consider how the technique can be used
to implement five basic elements required to build a quantum
computer, and then to discuss whether these techniques can be
scaled up for use in computers with large numbers of qubits.  This
approach is not appropriate for NMR, as it is already clear that
all the five basic elements can be constructed; indeed they have
all been experimentally demonstrated.  Thus, the only topic
remaining to be addressed is that of scaling.  It is, however,
useful to consider each of the five basic elements in turn, and
discuss their impact upon the practicality of building a large NMR
quantum computer.

The sections below are largely concerned with what might be called
``conventional'' liquid state NMR quantum computers, by which I
mean computers implemented using standard techniques from liquid
state NMR with spin-\half\ nuclei.  I will, however, briefly
discuss some relatively simple techniques, such as the use of
optical pumping or the use of liquid crystal solvents, which may
allow small extensions in the range accessible to NMR quantum
computers without greatly altering the underlying physics.  I will
not consider solid-state NMR: although ultimately based on the
same physical interactions the solid state NMR Hamiltonian
\cite{Abragam:1961, Slichter:1990, Schmidt-Rohr:1994} is
\emph{much} more complicated than the liquid state form, and
liquid and solid state NMR form two largely separate sub-fields.
While some solid state NMR experiments could be considered as
implementations of quantum simulations (\emph{e.g.},
\cite{Zhang:1998}), they have not as yet been used to build
general purpose quantum computers. Similarly, while nuclei with
spin quantum numbers greater than \half\ (quadrupolar nuclei) are
of some theoretical interest \cite{Kessel:1999, Kessel:2000},
their use raises considerable experimental difficulties.  Finally
I will largely ignore proposed systems which use single isolated
atomic nuclei in solid state devices, such as the proposal due to
Kane \cite{Kane:1998, DiVincenzo:1998}; although such systems are
ultimately based upon NMR, they differ from current liquid state
NMR implementations so profoundly that it is difficult to draw
detailed parallels.

\section{Initialization}
Initialization is the process of placing a quantum computer in
some well defined initial state, typically $\ket{\mathbf{0}}=
\ket{000\dots 0}$, prior to beginning the computation. In
principle any initial state is as good as any other, but in
practice \ket{\mathbf{0}} is the most widely chosen, both because
it corresponds to the traditional starting point of many quantum
algorithms and because it often corresponds to the system's
energetic ground state.  This is indeed the case in NMR quantum
computation, where the computational basis corresponds with the
natural experimental basis.

As initialization requires that the quantum computer be placed in
the state \ket{\mathbf{0}}, independent of its state before the
beginning of the initialization process, it is clear that it
cannot be achieved by any unitary process; thus initialization
schemes must be quite different from quantum logic gates.  As the
desired initial state is usually an energetic ground state,
initialization is typically achieved by cooling. This is not a
practical approach within NMR, as the energy gaps involved are
tiny compared with the Boltzmann factor at room temperature.

The energy gap between nuclear spin levels in NMR experiments is
principally determined by the Zeeman interaction between the
nucleus and the applied magnetic field (with the exception of the
quadrupolar interaction, which does not occur in spin-\half\
nuclei, all other nuclear spin interactions are small compared
with the main Zeeman interaction).  The Zeeman splitting, $\Delta
E=h\nu=\hbar\gamma B$, is usually described in terms of the
corresponding Larmor frequency, $\nu$, and is proportional to the
magnetic field strength, $B$, and the gyromagnetic ratio,
$\gamma$, which is an intrinsic property of the nucleus.  For the
magnetic field strengths typically used in NMR experiments
($2.3$--$21.1\,\text{T}$), the Larmor frequency of \nuc{1}{H}
nuclei lies in the range $100$--$900\,\text{MHz}$, corresponding
to an energy of $0.4$--$3.7\,\mu\text{eV}$.  For all other nuclei
(with the exception of the radioactive nucleus tritium,
\nuc{3}{H}), $\gamma$ is lower than for \nuc{1}{H}, with a
corresponding reduction in $\Delta E$.

These energies are much smaller than the Boltzmann energy at room
temperature (about $25\,\text{meV}$), and so at thermal
equilibrium the excess population in the lower Zeeman level is
tiny, less than one part in $10^4$.  For this reason conventional
liquid state NMR was long ruled out as a practical technology for
quantum computation.  In 1996, however, it was realised that it is
not strictly necessary to start quantum computations from a pure
state: a pseudo-pure, or effective pure state will suffice
\cite{Cory:1996}.

\subsection{Pseudo-pure states}
A pseudo-pure state is a mixed quantum state corresponding to a
mixture of the desired quantum state, \ket{\psi} and the maximally
mixed state, $\mathbf{1}/N$, where $N=2^n$ is the dimension of the
Hilbert space describing the system. To perform a quantum
computation it suffices to form the pseudo-pure ground state,
\begin{equation}
\rho=(1-\epsilon)\mathbf{1}/N+\epsilon\ket{\mathbf{0}}\bra{\mathbf{0}}.
\label{eq:psinit}
\end{equation}
An otherwise error-free quantum computer which begins a
computation in such a state will end up in the state
\begin{equation}
\rho'=(1-\epsilon)\mathbf{1}/N+\epsilon\ket{\mathbf{\psi}}\bra{\mathbf{\psi}}
\label{eq:psfinal}
\end{equation}
where \ket{\psi} is the result of the computation (assumed for the
moment to be an eigenstate in the natural basis). This result may
be immediately deduced by noting that any quantum computation
corresponds to some unitary evolution of the quantum state; such
evolutions are linear and have no effect upon $\mathbf{1}$.

A quantum computer in this mixed state will return the correct
answer with probability $\epsilon +(1-\epsilon)/N$, and a wrong
answer with probability $(N-1)(1-\epsilon)/N$ (note that the
maximally mixed state itself contains a fraction $1/N$
corresponding to the ground state).  Clearly it is possible to
determine the desired answer by statistical analysis of the
results of a sufficiently large number of repetitions of the
computation, where the number of repetitions required depends on
$\epsilon$.  Equivalently it is possible to use an ensemble of
quantum computers, and determine the ensemble averaged result of
the computation; as long as the ensemble is sufficiently large
this process will unambiguously point to the desired answer.

This latter approach is precisely that adopted for NMR quantum
computation, and indeed for conventional NMR spectroscopy.  NMR
transition frequencies are so low that it is essentially
impossible to detect a single transition, and so it is necessary
to use macroscopic samples, typically containing about $10^{17}$
molecules, with an excess population of about $10^{13}$ nuclei in
the low energy state.  It might seem that the small signal from
the excess population would be swamped by a huge background, but
this is not the case as the maximally mixed state gives rise to no
overall signal.  This is because the signals from different
components in the maximally mixed state cancel each other out; the
operators corresponding to the effective observables in NMR
spectra are all traceless.  Thus the NMR signal arises entirely
from the small excess population, and the signal from an NMR
quantum computer in the pseudo pure state, equation
(\ref{eq:psfinal}), is identical to that from one in the pure
state \ket\psi\bra\psi, except that the signal strength is reduced
by a factor of $\epsilon$.

\subsection{Assembling pseudo-pure states}
While pseudo-pure states offer the theoretical possibility of
performing quantum computations with mixed states, this approach
is useful only if some practical procedure for assembling such
states can be devised.  For the simplest possible quantum
computer, comprising a single qubit, the process is trivial, as
the thermal equilibrium density matrix has exactly the desired
form, but with larger systems the situation is more complicated.
For a system of $n$ spin-\half\ nuclei, all of the same nuclear
species (a homonuclear spin system), the $2^n$ eigenstates will be
distributed across an evenly spaced ladder of $n+1$ groups of
energy levels, with the number of (nearly degenerate) states
within each group given by Pascal's triangle; the population of
each state will be determined by the Boltzmann equation.  If the
system contains several different nuclear species (a heteronuclear
system), the situation is similar but slightly more complicated.
Normally NMR experiments are conducted in the high temperature
limit (see below), and so the pattern of excess populations will
simply be proportional to the energy of each state.

Assembling a pseudo-pure state from such a complex mixture might
seem difficult, but it is in fact a fairly conventional problem
from the viewpoint of NMR \cite{Cory:1996}. The apparent problem
is that reaching even a pseudo-pure state requires a non-unitary
process, and the most obvious such process (relaxation to thermal
equilibrium) leads to a state which is neither the desired state,
nor unitarily related to it. Thus some additional non-unitary step
is required.

In fact there are two elements commonly used in NMR pulse
sequences with non-unitary effects: magnetic field gradients and
phase cycling.  The first of these relies on the fact that the
sample forms a macroscopic ensemble; by applying Hamiltonians
which vary over the sample the \emph{ensemble averaged} evolution
can be non-unitary.  This is most commonly achieved by momentarily
destroying the spatial homogeneity of the main magnetic field (a
$B_0$ field gradient pulse \cite{Keeler:1994}), but similar
effects can be achieved using spatially inhomogeneous RF fields (a
$B_1$ field gradient pulse \cite{Zhang:1995}).  The second
approach relies on combining the results of several subtly
different NMR experiments by post-processing; as this is done by
classical methods, such processing is not confined to unitary
transformations.  In conventional NMR this is referred to as phase
cycling \cite{Bodenhausen:1984}, and plays a central role in many
experiments, although for many purposes it has now been replaced
by the use of gradient techniques.

Both of these approaches have been used to assemble pseudo-pure
states.  The original approach of Cory \emph{et al.}
\cite{Cory:1996}, based on magnetic field gradients, is in many
ways the most satisfying, but an alternative ``temporal
averaging'' scheme based on phase cycling \cite{Knill:1998} has
also proved extremely popular. Recently Knill \emph{et al.} have
described a simple and general approach \cite{Knill:1999} for
building pseudo-pure states in a system of any size; their scheme
may be used with either gradient or phase cycling techniques.

In addition to these schemes, another quite different approach,
logical labeling, was suggested very early on by Gershenfeld and
Chuang \cite{Gershenfeld:1997}.  Their approach is based on the
observation that while the thermal equilibrium spin density matrix
for an $n$ spin system ($n>1$) does not have the desired form,
equation (\ref{eq:psinit}), subsets of energy levels can be chosen
which \emph{do} have the correct pattern of populations; the
computation is then performed within this subset of states.
Although theoretically elegant this scheme appears to be more
complex than the other approaches, and only two experimental
implementations have been reported \cite{Vandersypen:1999,
Dorai:1999}.

\subsection{Scaling the system up}
While pseudo-pure states provide a practical approach for building
small NMR quantum computers it is not possible to simply scale
this approach up to larger systems.  The basic problem is the size
of $\epsilon$, or rather the manner in which $\epsilon$ scales
with $n$, the number of spins in the system.

The exact value of $\epsilon$ will vary with details of the scheme
used to prepare the pseudo-pure state, and so it is more useful to
consider an upper bound, which gives the maximum amount of
pseudo-pure state which can possibly be extracted from thermal
equilibrium.  This is equal to the population difference between
the lowest and highest energy states \cite{Warren:1997}, which for
an $n$ spin homonuclear spin system is given by
\begin{equation}
\epsilon=\frac{2\sinh(nh\nu/2kT)}{2^{n}\cosh^n(h\nu/2kT)}
\end{equation}
where $h\nu$ is the Zeeman splitting.  In the high temperature
limit ($h\nu\ll kT$) this expression simplifies to
\begin{equation}
\epsilon\approx\frac{nh\nu/kT}{2^n}.
\end{equation}
Thus, in the high temperature limit the amount of pseudo-pure
state which can be obtained decreases \emph{exponentially} with
the size of the spin system.  To overcome this it is necessary to
use an exponentially large sample, or some equivalent approach
such as repeating the experiment an exponentially large number of
times.  It therefore seems that the pseudo-pure state approach
does not scale.

In passing it should be noted that this is by no means a feature
unique to NMR quantum computation: it occurs for any ensemble
quantum computation scheme in the high temperature limit.
Physically this is because such a system has $2^n$ levels, and the
population deviations must be distributed among them; hence the
excess population in any one state is inevitably exponentially
small.

The exponentially small size of $\epsilon$ has caused some authors
to dismiss NMR quantum computation as a practical approach; while
this point of view has merit it may, as discussed below, be over
hasty. More recently, it has been suggested \cite{Braunstein:1998}
that NMR might not be quantum mechanical at all!  As NMR
experiments are conducted in the high temperature limit, the
density matrix is close to a maximally mixed state, and such high
temperature states can always be decomposed as a mixture of
product states (that is, states containing no entanglement between
different nuclei). As NMR states can be described without invoking
entanglement, they can therefore be modeled classically, although
the classical models involved may be somewhat contrived.  While
this conclusion is clearly correct it has proved difficult to
develop classical models which fully describe NMR experiments
\cite{Schack:1999}, and the real significance of these
observations remains unclear.

\subsection{High fields and low temperatures}
As the problem with using pseudo-pure states arises from operating
in the high temperature limit, the most obvious solution is to use
either low temperatures or high fields, so that this limit no
longer applies.  Unfortunately NMR lies so far into the high
temperature regime that this approach is unlikely to lead to
success.

The critical fields and temperatures required are given by
$\hbar\gamma B \sim kT$; for \nuc{1}{H} nuclei and a magnetic
field strength of $21.1\,\text{T}$ (the largest NMR magnet
currently available) this corresponds to $T\sim 0.043\,\text{K}$.
Reaching such temperatures is possible, but clearly the sample
will no longer remain in the liquid state!  At such low
temperatures only solid state NMR is possible.  Alternatively if
the sample is held at room temperature then a magnet with a field
strength $B\sim 150000\,\text{T}$ will be required; this lies far
beyond anything which is likely to be achieved in the foreseeable
future.

The arguments above do not entirely rule out the possibility that
some combination of high fields and low temperatures might one day
be used to achieve reasonable polarizations, and thus interesting
pseudo-pure states (in particular, it may be possible to generate
high polarizations under one set of conditions and then observe
them in another), but in the short term this does not seem a
particularly sensible approach.  There are, however, a wide range
of alternatives.

\subsection{Optical pumping}
A more subtle approach to increasing spin polarization is to use
techniques such as optical pumping; this has the effect of
decreasing the apparent temperature of the spin system without
affecting the rest of the sample.  Optical pumping techniques are
in fact quite widely used within NMR, but it is not yet clear
whether they can be usefully applied within NMR quantum
computation.

The best known optical pumping process within NMR is the
spin-exchange optical pumping of noble gas nuclei
\cite{Walker:1997}, most notably the spin-\half\ nuclei
\nuc{3}{He} and \nuc{129}{Xe} and the quadrupolar nucleus
\nuc{131}{Xe}.  The process involves conventional optical pumping
of the electron spin states of alkali metal atoms, followed by
spin exchange in binary collision pairs (He) or van der Waals
complexes (Xe).  The resulting highly polarized noble gases have
been used in a variety of NMR experiments, including NMR imaging
\cite{Albert:1994}.  Unfortunately noble gases are unsuitable for
constructing NMR quantum computers as they consist of isolated
atoms, that is systems containing only a single spin. It is in
principle possible to transfer the polarization from the noble gas
to more interesting species using a variety of cross polarization
techniques \cite{Gaede:1995, Navon:1996, Pietrass:1999}, although
it has so far proved difficult to obtain high transfer efficiency
except when transferring polarization to surface species in
microporous materials.

The second common form of optical pumping in NMR is quite
different: optical pumping in bulk semiconductors such as Si, GaAs
and InP \cite{Lampel:1968}.  This approach, which is related to
the DNP schemes described below, is confined to solid state
systems, and it is difficult to see how it might be used to
improve current liquid state implementations of NMR quantum
computers.

\subsection{Other approaches}
There are many other techniques which can be used to increase the
initial polarization in NMR experiments: the low sensitivity of
NMR is perhaps its biggest drawback in conventional spectroscopic
studies, and seeking to improve sensitivity is a common research
topic.

Perhaps the most important technique for sensitivity enhancement
is the nuclear Overhauser effect (NOE), which arises from the
correlated relaxation of two or more nuclei.  If the polarization
of one nucleus is perturbed from its equilibrium value cross
relaxation will transfer some of this perturbation to other nearby
nuclei.  This technique is widely used, both to enhance the
polarization of low sensitivity nuclei, and to probe internuclear
distances \cite{Neuhaus:1989}.  The maximum polarization gain
which can be achieved, however, is proportional to
$\gamma_S/\gamma_I$, where $\gamma_S$ and $\gamma_I$ are the
gyromagnetic ratios of the sensitive and insensitive nuclei
respectively, and so this method cannot be used to increase the
polarization of \nuc{1}{H}, which has the highest gyromagnetic
ratio among stable nuclei.

A better approach for \nuc{1}{H} nuclei is to use the original
Overhauser effect\cite{Overhauser:1953, Overhauser:1953b}, which
transfers polarization from electrons to nuclei.  Initially
greeted with scepticism \cite{Abragam:1989}, Overhauser's
theoretical predictions were confirmed by Carver and Slichter
\cite{Carver:1953}, who demonstrated huge Overhauser enhancements
in the spectrum of metallic \nuc{7}{Li}.  The Overhauser effect,
and related phenomena collectively known as Dynamic Nuclear
Polarization or DNP \cite{Hall:1997, Wind:1985,
Muller-Warmutht:1983}, can be used to generate quite large
polarization enhancements in a range of solid systems containing
unpaired electrons, and when combined with optical techniques for
pumping the electron polarization \cite{Lampel:1968, Patel:1999}
dramatic enhancements can be observed \cite{Iinuma:2000}. However
the technique performs poorly in the liquid state.

Another technique which gives enhanced polarizations is Chemically
Induced Dynamic Nuclear Polarization, or CIDNP \cite{Hore:1993}.
Despite the name CIDNP is unrelated to DNP; instead the
non-equilibrium polarizations arise from a spin sorting mechanism
which takes place during chemical reactions. This intriguing
effect has proved a powerful tool for investigating a range of
biomolecular systems \cite{Hore:1993, Lyon:1999}, but the
polarizations achievable are unlikely to be useful for quantum
computation.

Yet another possible approach is a family of experiments using
\emph{para}-hydrogen induced polarization, or PHIP
\cite{Natterer:1997}. When hydrogen molecules are cooled into
their rotational ground state, the Pauli principle dictates that
the nuclear spin wavefunction of the two \nuc{1}{H} nuclei must be
antisymmetric, and so the two nuclei must have \emph{opposite}
spin states.  If \emph{para}-hydrogen is used in an addition
reaction, for example adding $\text{H}_2$ across a carbon--carbon
double bond, then the product of the reaction will also have
non-equilibrium spin states; these can be converted into a greatly
enhanced polarization by conventional NMR pulse sequences. As
\emph{para}-hydrogen has a pure nuclear spin state it should in
principle be possible to produce completely polarized molecules;
in practice the enhancements are usually somewhat smaller
\cite{Natterer:1997}.

Unlike many of the schemes discussed above, PHIP works well in the
liquid state using fairly ``normal'' organic molecules, and so may
prove useful in NMR quantum computation.  Unfortunately the scheme
only allows the production of molecules which are highly polarized
at two sites, and while it is in principle possible to use two or
more addition reactions within the same molecule to produce
polarization at four or more sites creating an entire spin system
by PHIP addition reactions does not seem a plausible process.
However, as discussed below, in some cases it may suffice to
produce high polarizations at a single spin; in this case PHIP may
indeed turn out to be a useful approach.

\subsection{Reinitialization}
One important point, frequently neglected in discussions of this
kind, is that the methods described above are essentially methods
for initializing an entire quantum computer at the start of a
computation.  They do not obviously permit the selective
reinitialization of individual qubits in the middle of a
computation.  This makes it difficult to implement effective error
correction schemes, as discussed in section \ref{sec:errors}
below.

\subsection{Computational solutions}
In addition to the physical approaches outlined above, two
computational approaches might allow the problem of low spin
polarization to be bypassed.

The first approach, due to Schulman and Vazirani
\cite{Schulman:1998}, uses computational methods to purify mixed
states.  Their scheme works by concentrating the polarization from
a large number of weakly polarized spins into a small number of
spins which become strongly polarized; alternatively the
calculation can be though of as occurring inside a small
effectively pure subspace within the large spin space.
Unfortunately the size of the subspace which may be extracted is
$O(\epsilon^2 n)$, and so an $n$ spin quantum computer requires a
spin system with $O(n/\epsilon^2)$ spins; for the values of
$\epsilon$ achievable by direct cooling the resulting overhead is
enormous.  If, however, the polarization can be substantially
increased by other means then this might provide a good method for
further purification.

Another recent theoretical discovery may allow all these problems
to be sidestepped, as it may be possible to perform many important
quantum computations without starting from a pure (or even
pseudo-pure) ground state.  It has been known for some time that
some quantum computations can be performed with a starting state
comprising one pure qubit together with a number of qubits in the
maximally mixed state \cite{Knill:1998b}; more recently it has
been shown that Shor's quantum factoring algorithm can be
performed in this way \cite{Parker:2000} (but see section
\ref{sec:readout} below). If this proves to be a useful approach
then it could simplify the problem of constructing an NMR quantum
computer, as single qubit pure states may be substantially easier
to reach than their multi qubit equivalents (for example, by using
\emph{para}-hydrogen based schemes).

\section{Gates}
In comparison with initialization, building quantum gates is a
relatively simple process for NMR quantum computers. Indeed, as
described above, many conventional NMR pulse sequences used  in
the molecular sciences can be viewed as sequences of quantum
gates.

Any quantum gate can be built out of one qubit and two qubit gates
\cite{Barenco:1995a}, and early work concentrated on building
these gates in two qubit NMR quantum computers (NMR systems
containing two spins). Unlike many other proposed implementations
of quantum computation, these gates cannot simply be transferred
to larger spin systems without modification, and there was some
initial concern that this transfer might prove difficult,
involving an exponential increase in the complexity of the gate
design. In fact, while the gates do need to be modified, the
transfer can be done fairly simply, and approaches are known for
doing this with only quadratic overhead.  More recently some
authors have moved on from simply imitating gates used in
``conventional'' quantum computers, and started designing gates
which harness the full power of the NMR Hamiltonian directly.

\subsection{One qubit gates}
One qubit gates are easy to implement in NMR as they correspond to
rotations of a single spin within its own subspace.  This is most
simply described using the Bloch vector model \cite{Hore:1995,
Bloch:1956} in which the state of a spin (or indeed any qubit) is
represented by a point on the surface of the Bloch sphere.  One
qubit gates correspond to rotations on this sphere, and can be
achieved using resonant radiofrequency (RF) fields; the power and
length of the RF pulse determines the rotation angle, while the
phase of the RF radiation corresponds with the rotation axis
within the $xy$ plane.  Rotations around other axes can be
achieved by composite rotations, implemented by applying several
RF pulses in sequence \cite{Freeman:1997}.

This description assumes that it is possible to apply RF pulses
selectively to individual qubits.  In more conventional proposals
for implementing quantum computation such qubit selection is
achieved using spatial localisation: each qubit is stored on some
physical object with a well defined location.  This approach is
not possible within NMR for three reasons.  Firstly, each qubit is
not stored on a single spin, but rather in an ensemble of spins
distributed throughout the sample.  Secondly, as the sample is in
the liquid state individual spins are in continual rapid motion.
Thirdly, the wavelength of RF radiation (around 1 metre) is huge
compared with the separation between spins, rendering conventional
spatial localisation impossible (although this could in principle
be overcome by using techniques from magnetic resonance imaging
(MRI) combined with enormous field gradients \cite{Callaghan:1993,
Zhang:1998b}).

Instead of using spatial selection, NMR quantum computation relies
on frequency selection to pick out individual qubits.  Different
nuclear species have different gyromagnetic ratios, and thus
different resonance frequencies; therefore RF pulses on resonance
with one spin will have little or no effect on other spins. With
two or more nuclei of the same type the situation is slightly more
complex, but even these will have different resonance frequencies
as different chemical environments will result in different
chemical shift (shielding) interactions \cite{Hore:1995}.  In this
case it is necessary to use low power selective RF pulses
(``soft'' pulses) \cite{Freeman:1997}, or their multipulse
equivalents, Dante sequences \cite{Morris:1978}, in order to
excite one spin while leaving other spins with only slightly
different resonance frequencies untouched (the excitation
bandwidth of an RF pulse depends on its power). This approach
works well in small spin systems, but will be difficult in large
systems as discussed below.

\subsection{Two qubit gates}
Two qubit gates are slightly more complex, as they require some
sort of interaction between pairs of spins.  In liquid state NMR
quantum computers this is provided by scalar spin--spin coupling
\cite{Hore:1995}.  The NMR Hamiltonian describing two weakly
coupled spins is
\begin{equation}
\mathcal{H}=\omega_I I_z + \omega_S S_z + \pi J_{IS} 2I_zS_z
\label{eq:HIS}
\end{equation}
where, following NMR practice, the spin operators are described
using product operators \cite{Sorensen:1983, Hore:2000}, and
energies are written as multiples of $\hbar$. The one spin angular
momentum operators $I_z$ and $S_z$ are simply scaled versions of
the Pauli matrices (the use of I and S to describe two spin
systems is traditional \cite{Solomon:1955}); $\omega_I$ and
$\omega_S$ are the resonance frequencies (Larmor frequencies) of
the two spins, and $J_{IS}$ is the scalar spin--spin coupling
(J-coupling), usually measured in Hertz.  Note that this
simplified Hamiltonian is only correct in the \emph{weak coupling
limit}, that is when $|2\pi J| \ll |\omega_I-\omega_S|$.

The most obvious way to implement a two qubit gate, such as the
\CNOT\ gate, is to use soft pulse techniques to selectively excite
one spin only when its neighbouring spin is in one of its two
eigenstates \cite{Barenco:1995b}; these two transitions have
different energies as they are split by spin--spin coupling. This
direct approach, picking out one transition from the multiplet
corresponding to transitions of a single spin, is in fact
identical to the old selective population transfer experiment
\cite{Pachler:1973}, and so this is the oldest method for building
quantum logic gates. More recently this approach has been used to
build two and three qubit gates directly \cite{Linden:1998a,
Dorai:1999, Arvind:1999}.

This direct approach is, however, relatively unpopular.  Instead,
most authors \cite{Cory:1996, Cory:1997, Gershenfeld:1997,
Jones:1998a, Chuang:1998a, Chuang:1998b, Jones:1998b, Jones:1998c,
Marx:1999, Jones:1999a, Vandersypen:1999, Cory:1998a, Cory:1998b,
Laflamme:1998, Nielsen:1998, Madi:1998, Jones:1998d, Linden:1999a,
Linden:1999b, Leung:1999, Linden:1999c, Price:1999a, Price:1999b,
Jones:1999b, Collins:1999, Jones:2000c} prefer to use multipulse
NMR techniques \cite{Ernst:1987, Goldman:1988, Munowitz:1988,
Emsley:1992, Freeman:1997} to sculpt the Hamiltonian, equation
(\ref{eq:HIS}), into a more suitable form. This is analogous to
the replacement of selective population transfer by the INEPT
pulse sequence \cite{Morris:1979}.  The basic idea behind
Hamiltonian sculpting is to use spin echoes \cite{Freeman:1997,
Hahn:1950, Bagguley:1992} to refocus specific interactions in the
Hamiltonian, thus creating an effective Hamiltonian obtained by
rescaling elements of the original Hamiltonian, or indeed
completely deleting them. While this approach to implementing two
qubit gates can be described in a variety of different ways, they
are all fundamentally equivalent. The central feature is the
implementation of a controlled phase gate \cite{Jones:1998d}, such
as
\begin{equation}
\bm{\phi}=
\begin{pmatrix}
1 & 0 & 0 & 0 \\ 0 & 1 & 0 & 0 \\ 0 &
0 & 1 & 0 \\ 0 & 0 & 0 & e^{i\phi}
\end{pmatrix}.
\label{eq:phi}
\end{equation}
This can be decomposed in product operator notation as
\begin{equation}
\bm{\phi}= \exp\left[-i\times\smhalf\phi\times\left( -(\smhalf
E)+I_z+S_z-2I_zS_z \right)\right].
\end{equation}
The first term ($\smhalf E$) can be ignored as it simply
corresponds to an (undetectable) global phase; the remaining three
terms can be achieved by sculpting the Hamiltonian, equation
(\ref{eq:HIS}), into the desired form.

More recently a third approach for implementing two qubit gates
has been suggested \cite{Jones:2000a}.  Once again this method
relies on the use of controlled phase gates, but the phase shifts
are generated not by using conventional dynamic phases, but
instead by geometric phases \cite{Shapere:1989}, such as Berry's
phase \cite{Berry:1984}.  Berry phases have been demonstrated in a
wide variety of systems \cite{Shapere:1989}, including NMR
\cite{Suter:1987, Goldman:1996} and the closely related technique
of NQR \cite{Tycko:1987, Appelt:1994, Jones:1995, Jones:1997}.
They can be used to implement controlled phase shift gates in NMR
systems \cite{Jones:2000a}, but it seems that this approach has
few advantages for NMR quantum computation over the more
conventional dynamic approach; the idea may, however, prove useful
in other systems \cite{Ekert:2000a}.

\subsection{Gate times}
It is not sufficient simply to show that a gate can in principle
be implemented; it is also important to consider how long it takes
to implement it.  As discussed in section \ref{sec:Decoherence}
below, what matters is not the absolute time taken, but how this
compares with the natural decoherence time. It is, therefore,
important to consider what factors limit the rate at which NMR
quantum, logic gates operate.

One qubit gates are not only simple to build, they are also rapid
to operate.  As single qubit gates correspond to rotations, a
reasonable measure is provided by the inverse of the time required
to perform a $2\pi$ rotation.  For a fully heteronuclear spin
system (that is, a spin system containing only one spin of any
given nuclear species) this rate is limited only by the available
RF power and the breakdown voltages of the RF coils; typical
values lie in the range $10$--$100\,\text{kHz}$.  For a
homonuclear spin system the situation is more complex as it is
necessary to use selective excitation.  In this case the gate rate
is constrained by the frequency difference between the resonances
which must be excited and those which must be left untouched. This
can be seen either by considering a Fourier spectral model of
excitation \cite{Ernst:1987}, or by analysis of the ``jump and
return'' pulse sequence \cite{Plateau:1982}. Thus in homonuclear
spin systems the one qubit gate rate is often below
$1000\,\text{Hz}$.

Two qubit gates are typically much slower.  If they are
implemented by direct selective excitation, then the situation is
the same as for one qubit gates in homonuclear systems, except
that the relevant frequency splitting is the scalar coupling
between the spins, $J_{IS}$.  A similar argument applies if the
gates are implemented using multiple pulse techniques: the
relevant rate is the inverse of the time required to achieve the
``antiphase'' condition, $1/2J_{IS}$. Scalar couplings are quite
variable, but they are frequently less than $10\,\text{Hz}$; thus
the gate times required for two qubit gates can be quite long.  In
our two qubit NMR quantum computer based on cytosine, a two qubit
gate takes about $70\,\text{ms}$. If implemented using geometric
phases \cite{Jones:2000a}, the time required is even greater.

In order to reduce these two qubit gate times it is necessary to
increase the size of the spin--spin coupling constant. In general
this is not possible, as $J_{IS}$ is fixed by the chemical system
chosen (it is, of course, possible to choose a new system with a
larger coupling constant, but this approach is obviously quite
limited). It is, however, possible to change the apparent size of
$J_{IS}$ by partially reintroducing dipolar couplings.  The scalar
coupling is in fact a quite small effect in comparison with the
principal spin--spin coupling interaction, dipolar coupling.  This
arises from the direct magnetic interaction between two magnetic
dipoles, and in the high field approximation \cite{Ernst:1987}
\begin{equation}
\mathcal{H}_{D}=\frac{\mu_0 \gamma_I \gamma_S
\hbar(1-3\cos^2\theta)}{8\pi r^3}\left(3 I_z S_z - I\cdot S
\right),
\end{equation}
where $r$ is the length of the internuclear vector $\bm{r}$, and
$\theta$ is the angle between $\bm{r}$ and the magnetic field. In
solid samples this results in coupling between one spin and all
its neighbours, but in liquids and solutions rapid motion causes
$\bm{r}$, and thus $\theta$, to fluctuate, so that
$\mathcal{H}_{D}$ is replaced by its isotropic average, which is
zero.  For this reason dipolar coupling has little direct effect
in liquid state spectra (it cannot be entirely neglected, as it is
one of the main sources of spin relaxation). The scalar coupling
is a correction to the dipolar coupling which arises from the
Fermi contact interaction: valence electrons can interact with two
or more nuclei and thus mediate an interaction between them. Like
dipolar coupling scalar coupling is anisotropic, but its isotropic
average is non-zero; thus the relatively small isotropic scalar
coupling is the dominant spin--spin interaction in the liquid
state.

Dipolar coupling is removed because isotropic tumbling reduces it
to its isotropic average.  If the motion is \emph{anisotropic}
then the average may become non-zero; this can occur if the
anisotropy of the molecular magnetic susceptibility causes
molecules to align slightly with the magnetic field, resulting in
small residual dipolar couplings. The effect is largest when the
magnetic susceptibility is high, and this approach has proved
helpful in investigating proteins bound to oligonucleotides
\cite{Tjandra:1997}.  Larger alignments, and thus larger residual
couplings, can be observed using liquid crystalline solvents or
cosolvents \cite{Bax:1997}, which align strongly in the magnetic
field and then themselves act to order other dissolved molecules.

This approach has been used \cite{Yannoni:1999} to implement
Grover's quantum search on a two qubit NMR quantum computer based
on chloroform dissolved in a liquid crystal solvent.  The
effective Hamiltonian in this system is
\begin{equation}
\mathcal{H}=\omega_I I_z + \omega_S S_z + \pi (J_{IS}+2D) 2I_zS_z
\label{eq:HISD}
\end{equation}
where $D$ is the residual dipolar coupling \cite{Yannoni:1999},
allowing the gate rate to be increased by a factor of $8$. It is
not yet clear how generally useful this approach will be.

\subsection{Two qubit gates in larger systems}
The description above is not only confined to one and two qubit
gates; it is in fact confined to such gates in two qubit
computers.  If these gates are to be used to build networks with
three or more qubits it is necessary to consider how they will
function in larger spin systems.

The situation for one qubit gates is relatively straightforward,
as these can still be implemented using selective RF pulses
(although the problem of frequency selection, discussed below,
becomes more serious).  For two qubit gates, however, a more
detailed analysis is necessary.  For a general $n$ spin system,
the Hamiltonian is
\begin{equation}
\mathcal{H}=\sum_{i} \omega_i I^i_z + \sum_{i<k} \pi J_{ik} 2I^i_z
I^k_z, \label{eq:HN}
\end{equation}
including $n$ Larmor frequency terms for each spin, and a total of
$n(n-1)/2$ spin--spin coupling terms, connecting every pair of
spins. In real systems some of the $J_{ik}$ coupling constants
will be so small that they can be neglected, but it is still
useful to consider the most general case.  Consider, for example,
a system of three spins, conventionally called $I$, $R$ and $S$;
in this case there will be three Larmor frequency terms, and three
spin--spin couplings, $J_{IR}$, $J_{IS}$, and $J_{RS}$.

It is in principle relatively simple to implement gates using the
direct method.  Each transition is split into four (under the
influence of two couplings); if selective excitation is applied at
just one of these four frequencies this corresponds to a doubly
controlled three qubit gate, such as the Toffoli gate.
Alternatively by exciting two transitions two qubit gates, such as
\CNOT, can be achieved.  In practice in large spin systems it will
become difficult to pick out the right set of transition
frequencies, and it is likely that some transitions will overlap.

The more common approach is to use Hamiltonian sculpting to
convert equation (\ref{eq:HN}) into the desired form.  Clearly
this requires refocusing the $n-2$ additional Larmor frequency
terms, as well as the $(n-1)(n-2)/2$ extra spin--spin couplings.
This can be achieved using spin echoes \cite{Linden:1999a,
Jones:1999b}. Unfortunately it is not possible to use just a
single echo to refocus all the interactions; instead it is
necessary to consider how the echo sequences interact with one
another.  The simplest method is to nest spin echoes applied to
each spin within one another, but this na\"{\i}ve approach
requires an exponentially large number of refocusing pulses.
Fortunately this problem can be sidestepped by using efficient
refocusing sequences \cite{Jones:1999b, Leung:1999b}, which allow
refocusing to be achieved with a quadratic overhead.

It is rare to find a large spin system where all the couplings
have significant size.  Coupling is a fairly local effect and so
large coupling constants are only found between spins that are
fairly close within the spin system; thus the coupling network is
described by a non-complete graph \cite{Jones:1999b}.  This
greatly simplifies the problem: not only does it reduce the number
of couplings which have to be refocused, but it also simplifies
the echo patterns needed to refocus the Larmor frequencies
\cite{Linden:1999c, Jones:1999b}. Thus the overhead required for
refocusing echoes is greatly reduced.

It might seem that it would not be possible to implement all the
desired logic gates in such a spin system, as some of the
requisite spin--spin couplings are missing.  In fact this is not a
problem, as long as every pair of spins is connected by some chain
of couplings.  Quantum \SWAP\ gates \cite{Madi:1998, Linden:1999b}
can be used to move quantum information along this chain; the
overhead imposed by these \SWAP\ gates is at worst linear in the
size of the spin system.  In most cases the advantages of using a
partially coupled spin system significantly outweigh the
disadvantages.

\subsection{Multiqubit gates}
As described above, NMR techniques can be used to implement
conventional one and two qubit gates, and thus any desired quantum
network.  This approach, imitating existing theoretical models of
quantum computers, was adopted by all the early papers on NMR
quantum computation.  It is not clear, however, that this is the
best approach: it might be more sensible to consider what types of
gates NMR systems are good at providing, and then seeing whether
these are valuable in quantum computation.  To take a simple
example the natural two qubit gate for NMR systems is the
controlled phase shift, not the \CNOT\ gate, and in many quantum
algorithms phase shifts are exactly what is needed. For example,
NMR quantum computers can readily implement Grover's quantum
search without using an ancilla qubit to convert oracle calls into
phase shifts \cite{Chuang:1998b, Jones:1998b, Jones:1998c}.

A more complex, but more interesting, example is the
implementation of multiqubit gates.  As the NMR Hamiltonian,
equation (\ref{eq:HN}), contains terms connecting multiple pairs
of qubits, it should be possible to use this Hamiltonian to build
certain multiqubit gates directly.  This was briefly discussed
above, when considering the direct implementation of Toffoli gates
in a multi-spin system, but Hamiltonian sculpting should allow the
approach to be used more widely. To date this idea has received
only brief analysis \cite{Price:1999b}, but this is likely to be a
productive source of gates in the next few years.

\subsection{The problem of selective excitation}
From the descriptions above it might seem that implementing
quantum logic gates in NMR quantum computers is essentially
solved, and that the solutions scale fairly well. This is almost,
but not quite, true as one major problem remains: selective
excitation.

All the techniques described above rest on the assumption that it
is possible to address individual qubits, so that interactions can
be refocused.  For NMR quantum computation (or indeed any NMR
pulse sequence) this is achieved using frequency selection, rather
than more conventional spatial localisation techniques, and this
approach is only possible if the NMR transitions all have well
separated frequencies; in particular it is simplest if the
separation between any pair of Larmor frequencies is much greater
than the width of the NMR multiplets, that is the sum of the
spin--spin couplings.  Note that in contrast with conventional NMR
experiments it is not sufficient simply to selectively excite one
spin; it is also essential that other spins remain
\emph{completely} unaffected.

With small spin systems this is fairly easy to achieve, but with
larger systems it can become quite difficult. For example, the
range of Larmor frequencies found for \nuc{1}{H} nuclei in simple
organic compounds, and working at a  \nuc{1}{H} frequency of about
$500\,\text{MHz}$ is only about $5000\,\text{Hz}$, and this
limited frequency range can soon ``fill up''.  Increasing the
number of spins not only increases the number of NMR multiplets,
but also increases the width of each multiplet by introducing more
spin--spin couplings.  With other nuclei the situation is similar,
although usually less serious (the \nuc{1}{H} frequency range is
unusually narrow).

A partial solution is provided by turning to heteronuclear spin
systems.  As the NMR frequencies of different nuclei are very
different, it is trivial to achieve nucleus selective, and thus
spin selective, excitation.  The comparative simplicity of
heteronuclear NMR quantum computation explains its early, and
enduring, popularity.  Unfortunately this approach can not be
continued indefinitely, as the number of suitable nuclei is small,
the obvious candidates being \nuc{1}{H}, \nuc{13}{C}, \nuc{15}{N},
\nuc{19}{F} and \nuc{31}{P}.

It seems likely that the problem of selective excitation will
prove a serious difficulty in constructing large NMR quantum
computers.  Although it is difficult to assess exactly what the
limit will be, it is notable that the largest number of spins of
one nuclear type used to date is six \nuc{1}{H} nuclei
\cite{Linden:1999c}; one other paper has described computations
involving four \nuc{13}{C} nuclei and three \nuc{1}{H} nuclei
\cite{Knill:1999}, while all other authors have used at most three
spins of any one nuclear type.  Assuming that it is practical to
address six spins of each of the five nuclei listed above, this
suggests a limit of around 30 qubits imposed by the problem of
selective excitation.  Actually designing and synthesising such a
spin system is another problem entirely.

\section{Decoherence}
\label{sec:Decoherence} In order to perform large quantum
computations it is essential that errors be kept under control. In
practice this means that the decoherence time must be very long in
comparison with the gate time, although methods of error
correction (see section \ref{sec:errors}) allow this criterion to
be slightly relaxed. The situation for NMR quantum computation
might appear extremely good, as systems with very long relaxation
times are known: for example, the spin--lattice relaxation time
($T_1$) of \nuc{129}{Xe} can be thousands of seconds
\cite{Haake:1998}.  As scalar spin--spin coupling constants can
reach hundreds of Hertz, and dipolar couplings can be even larger,
a na{\"\i}ve calculation suggests that it should be possible to
implement about $10^4$--$10^6$ gates before decoherence becomes a
serious problem.

In fact this calculation is meaningless for two reasons. First,
these extremely long relaxation times are always spin--lattice
relaxation times; the spin--spin relaxation times ($T_2$), which
provide a better measure of the decoherence rate, are much shorter
(typically below ten seconds).  Secondly, the relaxation times and
gate rates are taken from different spin systems, and it is not
possible to simultaneously achieve them in a single molecule. The
long relaxation times observed for \nuc{129}{Xe} gas arise
precisely because Xe gas atoms have only weak interactions with
their neighbours, and such systems are of little apparent use for
quantum computation as they provide no mechanism for logic gates.
As the scalar coupling is related to an underlying dipolar
coupling, and dipolar coupling is one of the two principal sources
of relaxation in spin-\half\ nuclei, spin systems suitable for
constructing NMR quantum computers inevitably possess shorter
relaxation times.

One further feature of decoherence in NMR quantum computation
deserves consideration.  As NMR systems are ensemble quantum
computers, the effect of decoherence is not simply to introduce
errors, as occurs with more conventional designs.  Rather these
errors must be averaged over the ensemble, and if, as is often the
case, these errors are fairly random in character, the overall
effect is that the error terms will largely cancel out.  Thus the
principal effect of decoherence is to reduce the apparent signal
strength.  This is clearly visible in some quantum counting
experiments \cite{Jones:1999a, Cummins:2000}, where decoherence
appears as an exponential decay in signal.

It is difficult to estimate a realistic limit to NMR quantum
computation arising from decoherence; however, current experiments
have been performed involving hundreds of logic gates
\cite{Jones:1999a, Cummins:2000}, and it seems likely that the
limit is about one thousand gates.

\section{Quantum error correction}
\label{sec:errors} Quantum error correction, and its companion
fault tolerant computation \cite{Shor:1995, Steane:1996,
DiVincenzo:1996, Preskill:1998, Steane:1998, Steane:1999}, play a
central role in considerations of the practicality of large
quantum computers: while small quantum computers may handle errors
fairly well the fragility of highly entangled states renders large
quantum computations highly vulnerable.  Error correction tackles
this by diagnosing these errors and correcting them, while fault
tolerant computation provides methods for minimizing the spread of
errors, and in particular reducing the impact of errors in error
correction schemes.  Simple examples of such schemes have been
implemented on NMR quantum computers \cite{Cory:1998b,
Leung:1999}.

The importance of these techniques can hardly be overstated; until
their discovery many authors believed that it would be completely
impossible to build a large quantum computer as it would be
hopelessly error prone. Unfortunately this benefit comes at a
price: error correction involves a substantial overhead as each
logical qubit must be encoded using many ancilla qubits. While
this overhead has been substantially reduced by new codes
\cite{Steane:1999} it is still at least an order of magnitude.
More importantly, however, this assumes that it is possible to
reuse ancilla qubits in order to repeatedly correct errors; in
turn this requires the ability to reinitialize ancilla qubits at
will.  With current NMR quantum computers this cannot be achieved,
and it is necessary to use a supply of fresh ancilla qubits at
each stage. In this case the overhead becomes so large as to be
completely impractical.

\section{Read-out}
\label{sec:readout}  Once a quantum computation has been performed
it is necessary to use some read-out scheme in order to extract
the result.  As NMR quantum computation is implemented using an
ensemble of spin systems, read-out involves ensemble measurements,
and thus expectation values.  This can have a number of profound
consequences.

In simple cases NMR read-out is little different from more
conventional schemes.  Suppose that a quantum computation ends
with the answer qubits in eigenstates.  In this case it is only
necessary to determine whether each qubit is in state \ket{0} or
\ket{1}, which is equivalent to determining the expectation value
of $\sigma_z$. This can be achieved either by exciting the
corresponding spin with an RF pulse and observing the phase of the
NMR signal \cite{Jones:1998a, Jones:1999a}, or by examining the
multiplet structure in the NMR spectrum of a neighbouring spin, or
by a combination of these techniques \cite{Chuang:1998a}.

A more interesting situation occurs when the algorithm ends with
one or more answer qubits in superposition states.  A conventional
quantum computer will return one of the corresponding answers at
random, while an ensemble quantum computer will return an ensemble
average over the set of all possible answers
\cite{Gershenfeld:1997, Jones:1999a}.  In most cases this result
is not particularly useful, and it is necessary to recast the
algorithm so that a single well defined result is obtained
\cite{Gershenfeld:1997, Jones:1999a}.  In some special cases,
however, ensemble measurements can be advantageous
\cite{Jones:1999a}.

A similar situation arises when NMR techniques are used to
implement phenomena such as quantum teleportation
\cite{Nielsen:1998, Bennett:1993}.  Traditional teleportation
schemes use strong measurements to project an unknown quantum
state into the Bell basis; classical results from these
measurements are then used to determine which of a set of unitary
transformations must be used to finish the protocol.  In NMR
teleportation, however, such projective measurements and classical
readout cannot be used.  Instead it is necessary to use
conditional evolution to perform the final unitary transformation
\cite{Nielsen:1998}.

This lack of projective measurements may also have consequences
for quantum computation using a single pure qubit
\cite{Knill:1998b, Parker:2000}.  The model of Knill and Laflamme
\cite{Knill:1998b} assumes only ensemble measurements, and so
accurately reflects the nature of NMR quantum computation, while
the work of Parker and Plenio \cite{Parker:2000} assumes that
projective measurements can be made.  It remains to be seen how
significant this difference actually is.

\section{Conclusions}
It is useful to draw some overall conclusions about the potential
usefulness of liquid state NMR as a technique for implementing
\emph{large} quantum computers.  For \emph{small} quantum
computers liquid state NMR techniques are well ahead of the
competition: indeed in most areas of experimental quantum
information processing there quite simply is no competition!
However, there are several serious difficulties with extending
this approach to large systems, and it seems unlikely that any
very large liquid state NMR quantum computer will ever be built.

The issue most commonly raised is initialization: the pseudo-pure
states used in NMR are far from pure, with typical polarizations
below $10^{-4}$, and, more seriously, an exponential fall off as
the size of the computer is increased.  This alone would appear to
limit liquid state NMR quantum computers to about $30$ qubits.

In fact this assessment is probably too pessimistic.  There are
many techniques for increasing signal strengths, and while none of
them offer an immediate solution, several have potential.
Furthermore, recent theoretical results suggest that pure states
may be less important than previously believed. If low
polarizations were the only difficulty preventing the construction
of a large scale NMR quantum computer, then it seems highly likely
that a solution would be found.  Unfortunately there are other,
more serious, problems.

Constructing quantum logic gates in small systems is easy, indeed
almost trivial; this simplicity explains the rapid initial
progress in NMR quantum computation.  Early concerns about a
potential exponential increase in the complexity of implementing
these gates have proved unfounded, with the discovery of methods
for implementing two qubit gates in multi-spin systems with (at
worst) quadratic overhead.  Little thought has, however, been paid
to the problem of selective excitation and the crowding of
frequency space; in my personal opinion this is likely to be the
first serious barrier to building NMR quantum computers with many
more than ten qubits.

The problem of decoherence is, of course, common to all potential
implementations of quantum computers.  The relatively long
decoherence times of nuclear spins, which appear to make them good
candidates as qubits, arise because they interact only weakly with
their environment.  Unfortunately these weak interactions are also
manifested as low gate rates, and so the ratio of decoherence time
to gate time is not as large as one might hope. Nevertheless NMR
decoherence processes are fairly well behaved, and are not likely
to prove an insuperable problem in systems of ten qubits.

The difficulty of performing practical quantum error correction
schemes, arising from the lack of selective reinitialization, is a
serious problem for building \emph{large} quantum computers; this,
however, is only likely to become an issue if all other
difficulties are solved.

The inability to perform projective measurements, and the
difficulty of dealing with ensemble averaged data, is probably not
a serious problem in its own right.  This property makes it
difficult to use NMR experiments to test fundamental questions in
quantum mechanics, but has few major implications for computation.
Indeed in some cases, such as when considering the effects of
decoherence, it can actually be an advantage.  Unlike some other
issues, however, this limitations appears to be inherent in the
liquid state NMR experiment, and it is not at all clear how it can
be bypassed.

Finally, it is interesting to compare current liquid state NMR
quantum computers with Kane's radically different, but ultimately
related, proposal for a solid state NMR quantum computer
\cite{Kane:1998}.  It is notable that Kane's proposal keeps many
of the advantages of NMR, but also manages to tackle some of the
most serious difficulties described above.  Firstly, Kane's
proposal uses external ``gates'' to modulate both the Larmor
frequencies and the spin--spin coupling constants of nuclei.  This
should remove all the difficulties with implementing quantum logic
gates.  Secondly, the proposal includes a scheme for making
projective measurements on single spins, thus also providing a
simple initialization scheme.

Even if, as seems likely, a large liquid state NMR quantum
computer is never built, quantum computation will still owe much
to NMR.  By providing the first working quantum computers, no
matter how small, NMR has reinvigorated the field.  Tricks long
known to NMR spectroscopists are now being applied in NMR quantum
computations, and many of these will have applications in other
technologies.  NMR quantum computation has had, and still has,
much to offer.

\section*{Acknowledgements}
I thank Mark Bowdrey, Holly Cummins, Patrick Hayden, Peter Hore,
Charles Lyon and Tanja Pietra{\ss} for helpful conversations.  I
am grateful to the Royal Society of London for a University
Research Fellowship.

\end{document}